# *Static friction at fractal interfaces*

Dorian A. H. Hanaor,  Yixiang Gan,  Itai Einav

*School of Civil Engineering, University of Sydney, NSW 2006, Australia*

**Abstract:**

Tribological phenomena are governed by combined effects of material properties, topology and surface-chemistry. We study the interplay of multiscale-surface-structures with molecular-scale interactions towards interpreting static frictional interactions at fractal interfaces. By spline-assisted-discretization we analyse asperity interactions in pairs of contacting fractal surface profiles. For elastically deforming asperities, force analysis reveals greater friction at surfaces exhibiting higher fractality, with increasing molecular-scale friction amplifying this trend. Increasing adhesive strength yields higher overall friction at surfaces of lower fractality owing to greater true-contact-area. In systems where adhesive-type interactions play an important role, such as those where cold-welded junctions form, friction is minimised at an intermediate value of surface profile fractality found here to be in the regime 1.3-1.5. Our results have implications for systems exhibiting evolving surface structures.

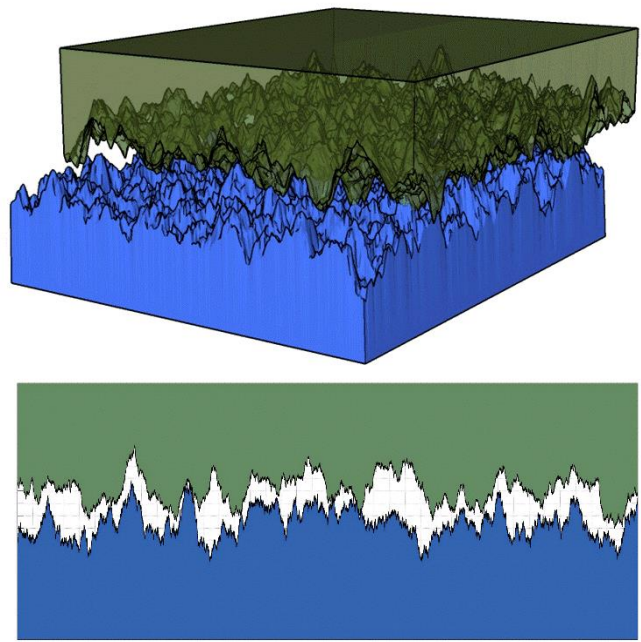

## 1. Introduction:

A meaningful micromechanical understanding of the origins of static friction and the ability to interpret its dependence on parameters of surface structure, surface chemistry, bulk material properties and environmental conditions is sought after in an extensive range of applications including granular materials [1, 2], electro mechanical devices[3], structural components[4] and across the broader field of applied mechanics.

In earlier approaches to the problem, static friction was considered to arise through the simple mechanical interactions of micro-scale asperities at contacting surfaces [5, 6]. The broadly observed linear dependence of frictional force on normal load of the Amontons Coulomb theory was often assumed to result from the presence of unseen surface features with a characteristic slope of $\alpha$ such that the coefficient of static friction follows $\mu_s = \tan\alpha$ [7]. Considering the significantly lower value of the true contact area relative to the apparent or nominal contact area, alternative approaches have considered friction arising from the shearing or debonding (cooperative or otherwise) of chemically bonded or welded junctions occurring at the regions of true contact [8-11]. A linear relationship between the total area of true contact and the applied load at an interface is ubiquitously found from numerical and experimental analyses and is indeed often utilised as a benchmark to ascertain the effectiveness of contact mechanics models[12-17]. This is understood to arise as the result of asperity hierarchies in elastically deforming surfaces[18], and following this rationale the typical linear Amontons-Coulomb behaviour can be said to arise through shearing or debonding of these regions, which are assumed to exhibit a constant shear strength [19]. A more inclusive representation of the origins of frictional interactions is given by understanding this as an integration of structural and molecular interactions across a range of scales as described by Bowden and Tabor [20-23].

While for most purposes a constitutive understanding of frictional behaviour as captured by the Amontons-Coulomb theory is sufficient, we frequently seek to gain a fundamental insight into the complex multi-scale and multi-physics interplay between surface structure, physico-chemical properties of materials and resulting frictional interactions [24]. This is of particular importance in multi-body systems such as granular materials, in small scale applications such as micro-electromechanical systems (MEMS), as well as in conditions of low loads where the resistance to shear, as has been often observed, may not exhibit linear dependence on normal forces [25, 26]. Importantly, couplings between structure and physico-chemical interactions at surfaces are of significance in systems exhibiting surface evolution and/or changing surface chemistry through changing environmental conditions or other time-dependant phenomena [27-29].

Molecular scale contributions to frictional interactions have been analysed in a range of materials and system configurations. Applying to the scale below that of measurable asperity structures, broadly this regime of effect can be divided into normal load dependant behaviour, which can be considered as atomic or molecular friction[10, 30-34], and contact area dependant resistance to shear generally considered as junction shear strength, contact bonding or adhesion[35-38]. These interactions have been studied using nanoscale experimental tools including friction force microscopy applied to atomistically flat surfaces[39, 40]. Data acquired through friction force microscopy is frequently interpreted using the Prandtl-Tomlinson model, which correlates surface structure with frictional behaviour and the occurrence of stick slip [41, 42]. Although highly significant in the field of tribology and atomic force microscopy (AFM), this model is limited to kinetic conditions where a localised body is traversing a rough surface, and is thus of limited applicability to the interpretation of frictional interactions in static conditions[43].

Various studies have investigated the dependence of frictional phenomena on parameters of mean surface roughness ($R_A$) and other surface roughness descriptors [44, 45]. Additionally, material properties and surface profile characteristics are often combined to give the indicative parameters such as the Plasticity Index, $\psi$, defined using parameters of hardness, asperity height distribution and asperity shape [46-48]. These studies have generally been constructed on the basis of a single distribution of asperity heights with assumed spherical features. However, naturally occurring surfaces tend to exhibit asperities at multiple scales in a fractal geometry exhibiting statistical self-similarity [49-52] and thus in recent years the fractal nature of surfaces has become a significant aspect in the field of experimental and computational surface analysis and contact mechanics [53-56]. The importance of considering surface fractality in contact mechanics can be explained by the tendency of first order roughness descriptors to be dominated by highest level features, while second order descriptors, such as mean slope or kurtosis, are dominated by the finest scales of surface features.

Using conventional finite element analysis (FEA) [38, 57, 58], Molecular Dynamics (MD) [33, 59] and discrete element methods[60, 61], challenges arise in the computationally efficient modelling of fractal surfaces and their contact mechanics, owing to the difficulties in capturing multiple scales in a single framework. Moreover, a significant majority of studies involving the contact mechanics of fractal surfaces have employed simplifications of rough to flat contact, limiting their applicability for studies towards static friction. In the present work we examine static friction occurring at interfaces of fractal surfaces in mutual contact as illustrated by Fig. 1,  using a method based on spline assisted asperity discretisation.

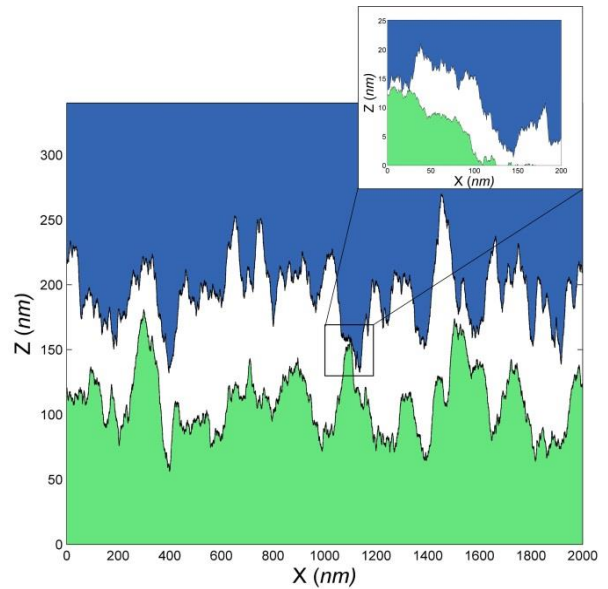

**Figure 1. Two approaching fractal surfaces showing a hierarchical structure typical of natural surfaces, as simulated in the present work.**

## 2.  Methods:

### 2.1. Generation of fractal surface profiles:

Static interactions between pairs of simulated fractal surfaces, representative of engineering surfaces, such as those shown in Figure 1, are considered where both surfaces are generated using the same value of fractality as described by the fractal dimension, $D_f$. To avoid the dominance of a small number of the highest surface features, rather than simulate a single pair of very long surfaces, simulations were carried out with repetition where two-dimensional fractal surface profiles of $1.5 \times 10^5$ (x,y) points were repeatedly generated (100 repetitions) and analysed with results averaged across the repetitions for each surface condition studied. With surface profiles of 40 microns in length, the finest features are thus equivalent to 2.67 Å, a figure which is towards the lower end of values typical of the separation between lattice planes in many crystalline materials.

In similarity to previously reported work, fractal profiles at each repetition $r$ (from $r$=1 to $r$=100) were generated through a method based on the Ausloos-Berman variant of the Weirstrass Mandelbrot function [55, 62, 63].

$$y_{(x)} = L^{(4-2D_f)} \ln \gamma \sum_{m=1}^{M} \sum_{n=1}^{n\max} \gamma^{2(n-1)(D_f-2)} \left\{ \cos \phi_{m,n,r} - \cos \left[ \frac{2\pi\gamma^{n-1}x}{L} \cos \left( -\pi \frac{m}{M} \right) + \phi_{m,n,r} \right] \right\}$$

(1)

following the present method is independent of the profile amplitude which is scaled to a selected height.

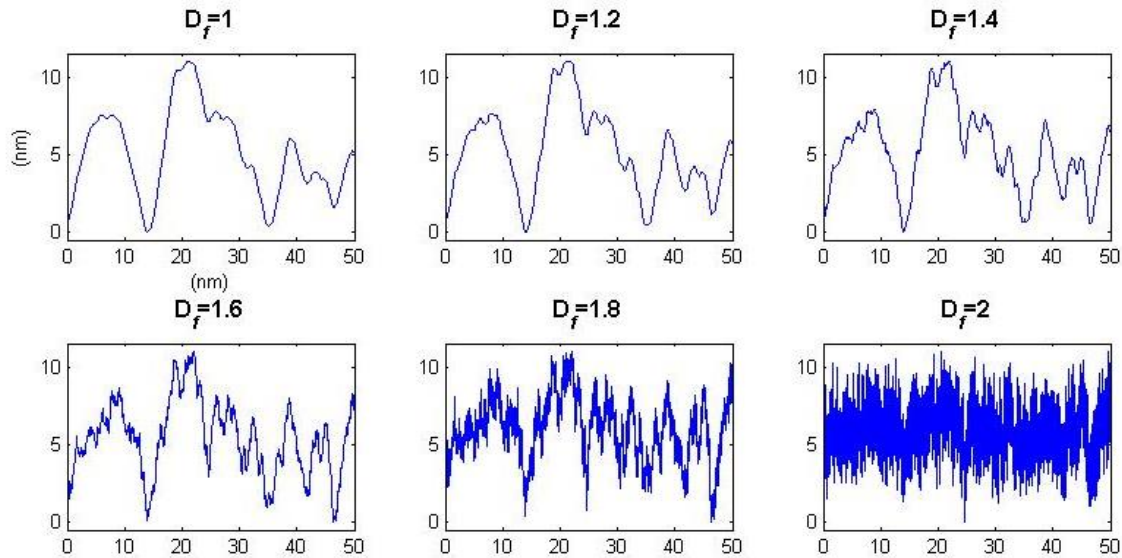

**Figure 2. Representative surface profiles over lengths of 5L, with constant rougness ($R_A$=1), and varying D values.**

The fractal dimension of surface profiles $D_f \in$ (1,2) relates to the scale variance of the surface structure as described by its surface roughness power spectrum [64]. Hence through methods applied here $D_f = 1$ corresponds to a smooth continuous quasi-random curve and $D_f = 2$ corresponds to a hypothetical area filling profile, within constraints of the simulation resolution. The effect of varying fractality is illustrated by representative profile sections shown in Figure 2. In the present work, for force evaluation we considered the $D_f$ values between 1 and 1.7 as this interval represents the range most relevant for real material surfaces, while the use of surface profiles exhibiting higher fractal dimensions would be only of theoretical interest. In contrast to the determination of fractal parameters from real surfaces, the simulation of surface profiles

As with previous work, a stochastic length parameter, L, has been included in Eq. 1 to account for higher level surface features, which are present with typical wavelength. In real surfaces the term L represents a characteristic macro-asperity spacing such as that which may arise from a granular structure. The effect of varying the parameter L is illustrated by the 1 micron profile sections shown in Figure 3. The profile used in contact simulations, with nominal lengths of 40 microns, were scaled in the y-direction to yield a consistent amplitude of asperities of 0.1 relative to the stochastic length parameter. In a separate set of simulations, the amplitude was varied to investigate effects of asperity aspect ratio. The parameter $\gamma$ represents the density of frequencies used to construct the fractal profile, which on the basis of reported methods is appropriately assigned as 1.5 [65-67]. The randomised phase angle $\phi$ is given by a uniform distribution of size $M$ x $n_{max}$ x $100$. $\phi_{m,n} = U(0,2\pi)$. In the present work $M$ and $n_{max}$ values of 40 were chosen as these were found to give sufficiently randomised surfaces.

which typically exhibit roughness features in this scale regime.

### 2.2. Spline assisted asperity discretization

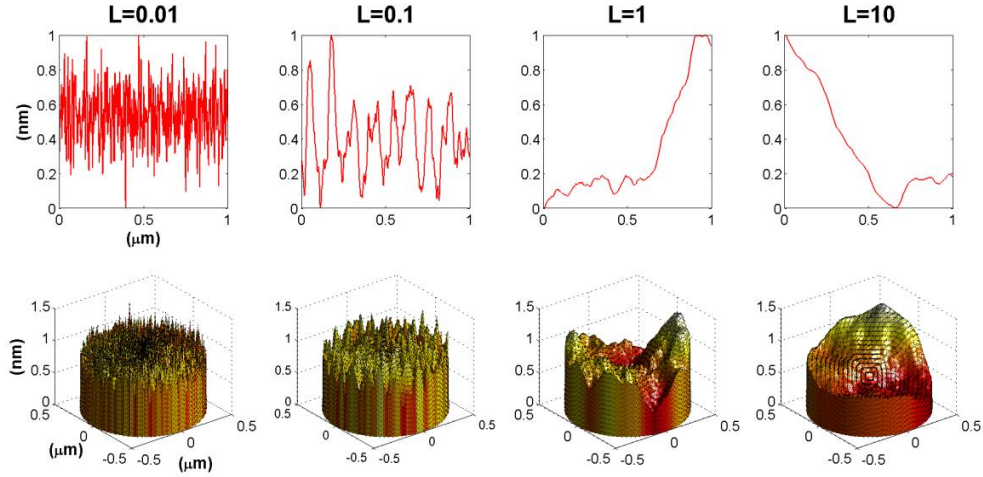

**Figure 3. Effect of the stochastic length parameter (L) in the simulation of 2D fractal profiles (upper) and equivalent 3D surfaces (lower).**

To represent macroscopically flat interfaces, repeatedly simulated pairs of fractal surface profiles were generated over a length of 40 L, to give on average 40 largest scale surface features per surface per repetition. To yield dimensional results (mN, μm) the value of L is assigned as 1 micron. Surfaces of varying fractality were studied over 100 repetitions. For each repeated contact scenario studied, two simulated surfaces, upper and lower, were generated with differing randomised $\phi$ sets. To yield macroscopically similar profiles comparable to the profiles illustrated in Figure 2, these sets of randomised phase angles $\phi$ were conserved for each repetition across the generation of 50 sets of surface pairs with increasing fractal dimension, varied from $D_f$=1.0 to 1.7. That is to say the $n^{th}$ repetition of surfaces generation is conducted with the same upper and lower $\phi$ sets regardless of the $D_f$ value. Surface profiles are assumed to be of unity thickness with micron dimensions used in order to show results pertinent to macroscopically flat engineering surfaces

Owing to the non-differentiability of fractal surfaces, surface normals and radii of curvature at discrete surface points on simulated profiles are extracted using a method of spline assisted asperity discretization (SAAD) as applied previously [68]. This allows the global contact problem to be treated as a series of local contact events. Following this method surface points are discretised using a cubic spline interpolation passing through all simulated points to describe surface features in terms of meaningful values of surface orientation and curvature radii, determined from the spline derived piecewise polynomial, $f(x)$ at individual points$(x_i, y_i)$

$$\vec{n}_i = \left[ \pm f'(x_i)\left(f'(x_i)^2 + 1\right)^{-0.5}, \mp \left(f'(x_i)^2 + 1\right)^{-0.5} \right]^T$$
(2)

$$R_i = \frac{\left(1 + f'(x_i)^2\right)^{3/2}}{f''(x_i)} \ , \quad \vec{O}_i = \vec{x}_i - R_i \, \vec{n}_i$$
(3)

Here $\vec{n}_i$ is the surface normal represented by a vector of unity magnitude the orientation of which varies depending on whether the surface is upper or lower in the contact event. $R_i$ is the local surface radius and assumes negative values for concave regions of the surfaces and tends towards infinity for a perfectly flat surface, while $\vec{O}_i$ represents the position vector of local sphere centres.

Following this discretisation, points involved in contact events are treated as Hertzian spheres where the relative positions of sphere centres for contacting asperities are utilised in order to compute the magnitude and orientation of localised normal and tangential forces as well as the areas of individual contact patches at active asperities. Forces and areas are then summed to yield global forces acting on the surface, which is assumed not to exhibit macroscopic flexure, as well as total true contact area. A schematic illustration is given in Figure 4.

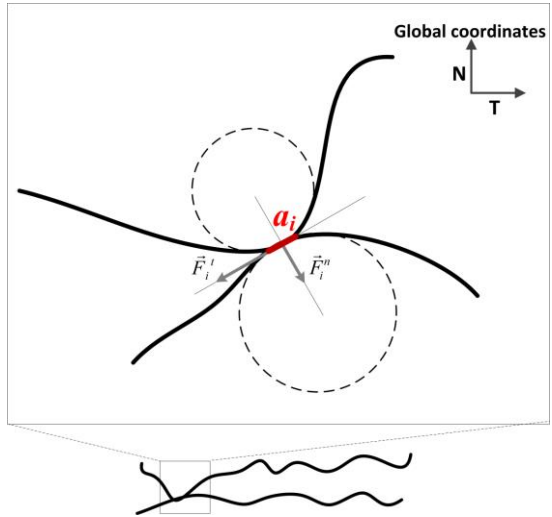

**Figure 4. Diagram of forces, normals and incremental displacements at a single contact point.**

### 2.3. Force evaluation

In the presently applied method we examine a static snapshot of discretised asperity interactions in normal and tangential orientations. Contact detection involves initially the identification of points satisfying the condition $y_i^u \leq y_i^l$, where superscripts $u$ and $l$ denote upper and lower surfaces. Secondly for each contact point a contact normal, in the form of a vector of unity magnitude $\vec{n}_i^c$, and a contact centre $c_i^c$ are evaluated.

$$\vec{n}_i^c = \frac{\vec{O}_i^u - \vec{O}_i^l}{\left|\vec{O}_i^u - \vec{O}_i^l\right|} = \frac{\vec{X}_i}{\left|\vec{X}_i\right|} \quad ,$$

$$\vec{c}_i^c = \frac{\vec{X}_i}{2} - \frac{R_i^u - R_i^l}{2}\,\vec{n}_i^c\,\mathrm{sgn}\left(R_i^u R_i^l\right)$$

$$(4)$$

Here $\vec{X}_i$ is the vector separation of the contacting sphere centres. It should be noted that the contact normals differ from the surface normals defined by SAAD.

Thirdly, as the grid of the simulated surface is of sufficiently high resolution to exhibit locally smooth topology, the number of actual contacting asperities is lower than the count of intruding grid points satisfying the aforementioned condition. For this reason, to avoid the force evaluation at single contacting asperities being represented by fluctuating numbers of spheres we exclude contacts between spheres on opposing surfaces occurring at points distant from their respective surface profile coordinates and thus avoid mesh dependence and discontinuities in force evaluation. Thus contact points are accepted only when their centres are located within the domain defined by

$$x_i^c \in \left[\, x_i - \frac{\Delta x}{2} \;\;,\;\; x_i + \frac{\Delta x}{2} \,\right]$$

$$(5)$$

where $x_i^c$ is the global $x$-axis coordinate of the contact centre and $\Delta x$ is the grid x-spacing of the surface-profiles. In the absence of this criterion, a single peak to peak contact point may be represented by a large number of spheres corresponding to distant points, and consequently yield an erroneously high local normal force.

At individual accepted contact points local normal forces and individual contact areas are resolved following the two dimensional Hertzian solution for elastic spheres [69, 70]. However, alternative solutions can readily be incorporated to accommodate divergent surface mechanics and material behaviour including asperity elasto-plasticity[71]. In the presently reported methodology an effective elastic modulus of $E^* = 10\ GPa$ is utilised in order to yield dimensional results relevant to engineering surfaces.

### 2.4. Friction evaluation

We aim to quantify static friction arising from interactions of asperities in simulated interfaces of fractal surfaces, and to gain insights into relationships between surface structure and friction. Additionally we incorporate parameters meaningfully representing molecular friction and adhesion to account for known physico-chemical interactions at material surfaces and examine the interplay between these parameters. In materials these interactions are governed by localised properties including surface crystallography, surface chemistry, the presence of adsorbates and electrostatic interactions.

The contact of atomistically smooth surfaces involves frictional interactions despite the absence of measureable asperities. These interactions, referred to varyingly as atomic, molecular, phononic or interface friction arise from intermolecular forces, electronic and van der Waals interactions at interfaces and are profoundly affected by the presence of adsorbed species [72-74]. Additionally, limitations to our ability to characterise asperity structures mean that surfaces may exhibit roughness features smaller than the scale which we are able to measure or meaningfully simulate. In the present study of rough to rough contact conditions these considerations are addressed by including a molecular / atomic friction coefficient, $\mu_0$, studied over the interval 0.1 to 1, to account for all interactions at scales below the scale of simulated asperities in the present work. It is assumed that this atomic friction coefficient is homogenous at all contact points regardless of height, orientation or contact length, although in natural surfaces variations in surface chemistry and crystallographic orientation would be expected to bring about inhomogenieties in this parameter. Following this the maximal locally tangential force that can be borne at an individual contact patch is given as

$$\vec{F}_i^t = \mu_0 \left| \vec{F}_i^n \right| \vec{t}_i^c \qquad (6)$$

where $\vec{t}_i^c$ is a unit vector in the local tangential orientation.

Finally the global coordinate system force balance is examined by summing the components of individual local forces at contact points in the global normal and tangential directions. In the global tangential direction we only sum forces corresponding to contact points where $F_i^{n,T} < 0$, that is to say where asperity interactions oppose an applied shear strain:

$$F_N = \sum \left( F_i^{n,N} + F_i^{t,N} \right), \quad F_T = \sum \left( F_i^{n,T} + F_i^{t,T} \right)$$

$$(7)$$

where $F_i^{n,N}$ and $F_i^{n,T}$ represent respectively the global normal and global tangential components of the local normal force acting at contact point *i*. Inherent to this methodology is the simplified assumption that through processes of deformation or micro-slip [75, 76] asperity contacts opposing shear reach their maximum local tangential force while asperities that do not oppose shear become unloaded.

### 2.5. Adhesion type interactions

The forces contributing to observed frictional phenomena across multiple scales arise from combined mechanical, electrostatic and molecular mechanisms. These have been investigated in recent years in a range of publications [59, 77-80]. In a constitutive approach at the finest scales frictional stress $\tau_f$ can be described by a load dependent component α (equivalent to $\mu_0$), and a material-interface-dependant adhesive shear stress $\tau_0$ such that $\tau_f = \tau_0 + \alpha P$ [81, 82]. With P being the contact pressure that is the sum of applied and capillary-induced components. The relative significance of the material and load dependant components is strongly influenced by the contact profile and fractality of surface structures.

Consequently for asperities involved in resistance to shear, we include in our force evaluation of frictional forces an adhesive component of varied significance, $F_i^A$. This is evaluated as $\vec{F}_i^A = \tau_0 A_i \vec{t}_i^c$, with overall friction evaluated as $F_T = \sum \left( F_i^{n,T} + F_i^{t,T} + F_i^{A,T} \right)$.

Similarly, normal forces are evaluated including the normal component of adhesive forces in the global normal orientation. In the present work we studied $\tau_0$ values of 500 KPa, 10 MPa, 100 MPa and 200 Mpa, to represent a range of natural adhesive interactions, such as those that might arise from can der Waals forces, micro-capillary forces or cold welded junctions of engineering alloys [80, 83-85]. Additionally we apply the current methods without the inclusion of $\tau_0$ and the contact area dependent forces that arise from this parameter.

For varied conditions of fractality ($D_f$), atomic friction ($\mu_0$) and adhesive shear strength ($\tau_0$) the macroscopic friction coefficient is evaluated in a straightforward manner by the overall ratio of $F_T$ to $F_N$ in the global coordinate system.

### 3. Results

#### 3.1. Frictional interactions of fractal surfaces

From the SAAD based analysis of contact events between pairs of amplitude-normalised fractal rough surfaces, conducted to a constant level of normal displacement, it is found that the evolution of true contact area with applied normal load closely follows a linear relationship, with some deviation at low normal loads. Results here are dimensional owing to the assumption of a profile of 1μm thickness and 40 μm in total length, and an effective modulus 10 GPa, however this is chosen only for illustrative purposes and non-dimensionalisation is readily possible. The dimensional value of true contact area is underestimated if simulation resolution is decreased, however this effect is negligible at the resolution level employed here studied here, which is sufficiently high to allow the representation of the scale regime of realistic macroscopically flat engineering surfaces (surfaces with features at length scales from $10^{-3}$m to $10^{-10}$m). The maximal $F_N$ values obtained differ with $D_f$ as the simulations were displacement driven. Overall the observed linear behaviour of true contact area is expected for hierarchical surface structures as predicted by Archard [18] and as confirmed in a range of experimental studies and by computational methods including boundary element method (BEM) [86] and FEA[57, 87, 88]. The results shown in Figure 5 were acquired from repeatedly simulated profiles with conditions of $\tau_0$=10 MPa and $\mu_0$=0.4, although these two parameters have a comparatively insignificant effect on the behaviour of true contact area and normal contact stiffness with applied normal load.

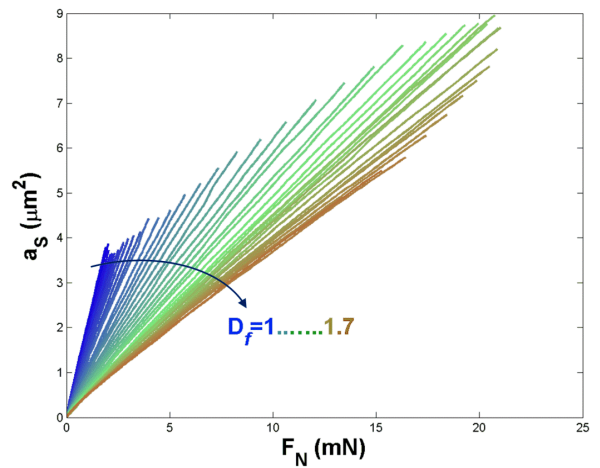

**Figure 5. Variation of true contact area with applied normal load for simulated surface profiles generated with $D_f$= 1 - 1.7.**

Following the methods described for the evaluation of macroscopically observed friction, the significance of loading conditions and surface structure are studied for a range of systems differentiated by parameters of their molecular scale friction coefficient ($\mu_0$) and adhesive shear strength at regions of true contact ($\tau_0$). As with all studies here, data points are established by averaging over the repeated generation of 100 surface profile pairs, with results shown in Figure 6. Although applied in a simplified manner in the present method, these two parameters are intrinsically linked to conditions of surface chemistry, temperature and environment and vary greatly with material type, surface orientation and environmental conditions. Here we examine $\mu_0$ values in the region 0.1-1.0, which covers the range of atomic friction coefficients typically reported for the study of atomically flat surfaces[89, 90]. Adhesive (load independent) interactions were studied by applying a shear strength $\tau_0$ in the regime of 500 KPa to 200MPa. As expected owing to the low value of true contact area relative to nominal contact, little difference was found between conditions of $\tau_0 \leq 1$ MPa, and conditions where adhesion type interactions are neglected i.e. $\tau_0 = 0$.

As higher surface slopes result greater normal forces at contact points, surfaces of greater fractality unsurprisingly exhibit a general tendency towards higher resistance to shear and a larger macroscopically observed friction coefficient shown on the right side of the plots in Figure 6. At regions of both high surface fractality and low applied normal load, the apparent macroscopic friction tends towards higher values following a similar pattern for all studied systems shown in the bottom right corners of the plots. It is worth noting that the results here show that for particular surface structures and applied loads we expect to encounter certain conditions where the macroscopic friction is lower than the molecular scale friction coefficient. That is to say the friction coefficient may be greater at a rough interface than at an atomically flat surface having the same material properties (bulk and surface). This can be explained by the fact that, unlike completely smooth surfaces, at rough interfaces, contact events are localised to a limited number of areas. In the present method, while the normal load $F_N$ is contributed to by forces at all asperities involved in the contact event, the maximal tangential force $F_T$ is contributed to only by asperity contacts oriented such that they resist shear. This result is supported by experimental results from various materials where the presence of increased roughness is often found to decrease the magnitude of frictional interactions within certain regimes under given loads, although it should be noted that this has been studied with respect to dynamic interactions [91, 92].

Conditions where shearing of cold welded asperity interfaces play a significant role, i.e. high $\tau_0$ values, yield an increase in overall friction across all levels of fractality and at all applied loads, however this trend is most evident towards low fractality and low applied normal load. Under such conditions overall friction is found to exhibit a minimum at intermediate fractal dimension values in the range ~1.3-1.5. Interestingly, this result demonstrates that for particular systems we may observe a non-monotonous variation of static friction with surface fractality. An increasingly high molecular friction coefficient naturally manifests in higher overall friction under all conditions with this being most prominent for surfaces with $D_f > 1.5$.

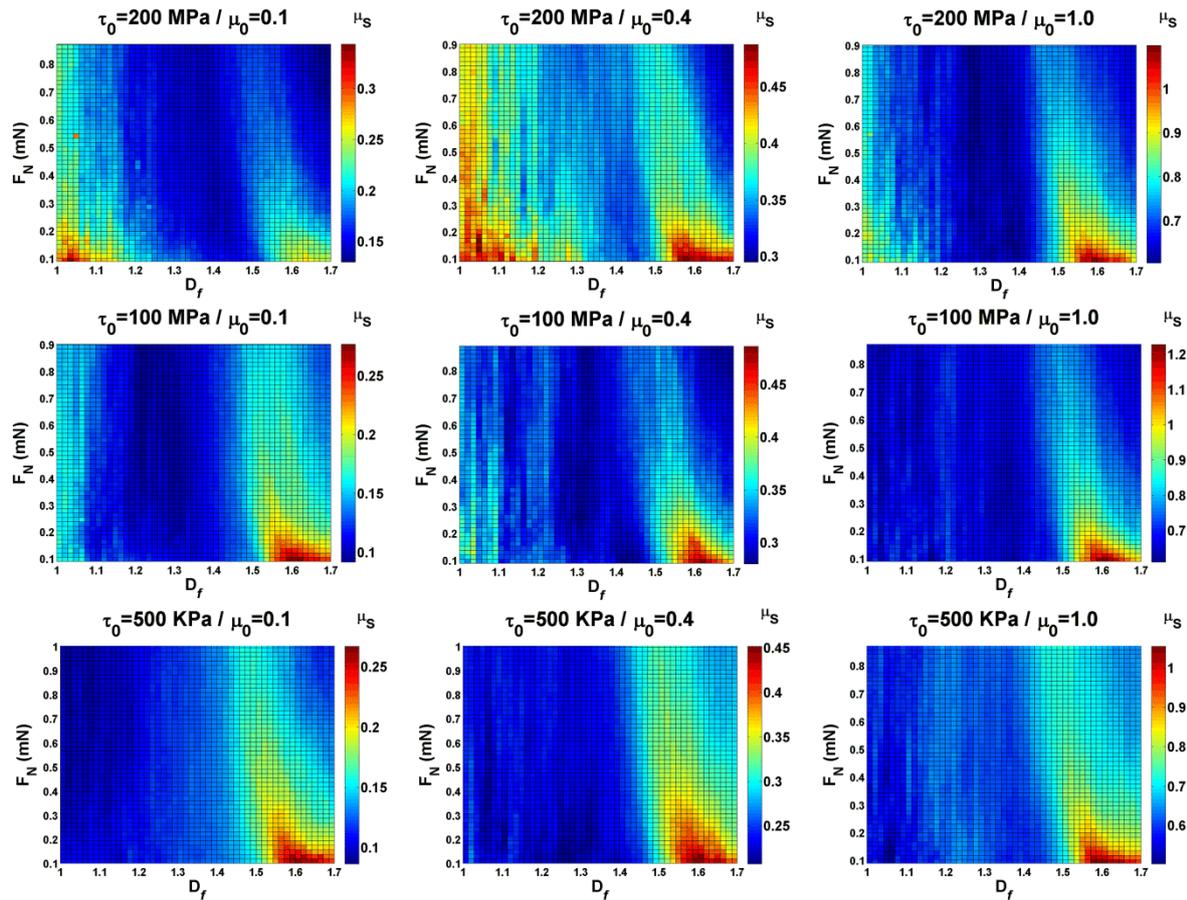

**Figure 6. Variation of macroscopic friction coefficient ($\mu_S$, colour bars) with fractal dimension and applied normal load for conditions of different adhesive strength ($\tau_0$) and atomic friction ($\mu_0$).**

### 3.2. Aspect ratio effects

To study the effects of asperity aspect ratio, the controlled amplitude was varied with respect to the stochastic length parameter L. This has the effect of governing the approximate aspect ratio of highest level asperities in terms of asperity height to asperity projection in the horizontal plane. In most engineering materials the aspect ratio is significantly less than 1 as asperity heights are generally significantly lower than the mean spacing (approximate wavelength) of highest level features. As with molecular scale interactions, the typical aspect ratio of asperities on macroscopically flat surfaces is highly dependent on the material type, with ductile materials showing low aspect ratios while ceramics and harder alloys exhibit higher asperity peaks relative to their horizontal size [93, 94].

The trend of higher static frictional strength towards regions of higher surface fractality is less prominent for surfaces with broader asperities, that is to say a lower amplitude relative to the value of L. This may be the result of lower surface slope values resulting in large number of contact points involved in frictional events. Moreover it can be seen that where adhesive interactions dominate the trend of higher friction at low surface fractality is more pronounced for flatter asperities, as the result of higher level of true contact area thus occurring. For surfaces dominated by contact bonding (higher adhesion) such those where static friction is assumed to arise from cold-welded junctions ($\tau_0$ =200 MPa), it is further worth noting that with increasing relative asperity amplitudes, the friction coefficient across all surface types and loading conditions exhibits a decreasing trend. This is the result of a decreasing true contact area at a given load, as the asperity aspect ratio increases, and the relative contribution of the adhesion to the macroscopic friction coefficient decreases. This is further illustrated in Figure 8, which shows the variation of true contact area with normal load for different sets of fractal surfaces at 3 different controlled profile amplitudes.

## 4. Discussion

We have utilised the SAAD method for the evaluation of contact mechanics and frictional interactions between pairs of multiscale rough surfaces. While this study employed profiles simulated using procedures for the generation of realistic fractal surfaces, the methodology applied could be used in conjunction with surface data acquired from AFM, surface profilometry or through other surface simulation algorithms. This approach involves rough-to-rough analysis in contrast to the majority of comparable work that most often involves rough to rigid flat simplifications for the study of similar interactions. The methods employed here have the notable advantage of facilitating the examination of the integrated effects of surface structure and surface chemistry on frictional interactions at rough to rough interfaces, in a manner that is computationally efficient relative to comparable FEA methods. Despite advantageous aspects of this model, there exist notable limitations of the presently applied approach. Importantly the present method examines a static snapshot for given structures and material parameters, and does not consider time dependant surface evolution, temperature effects on molecular scale interactions, plastic deformation of asperities and the discontinuously distributed tangential displacement of the surfaces as may occur through processes of deformation and microslip [33, 75, 76, 95, 96]. Furthermore, in studying mechanisms of bonding interactions (varied $\tau_0$) at regions of true contact, we assume these areas to be governed by Hertzian contact mechanics whereas a more suitable approach would be to consider cold-welded regions and Hertzian regions through separate frameworks [97].

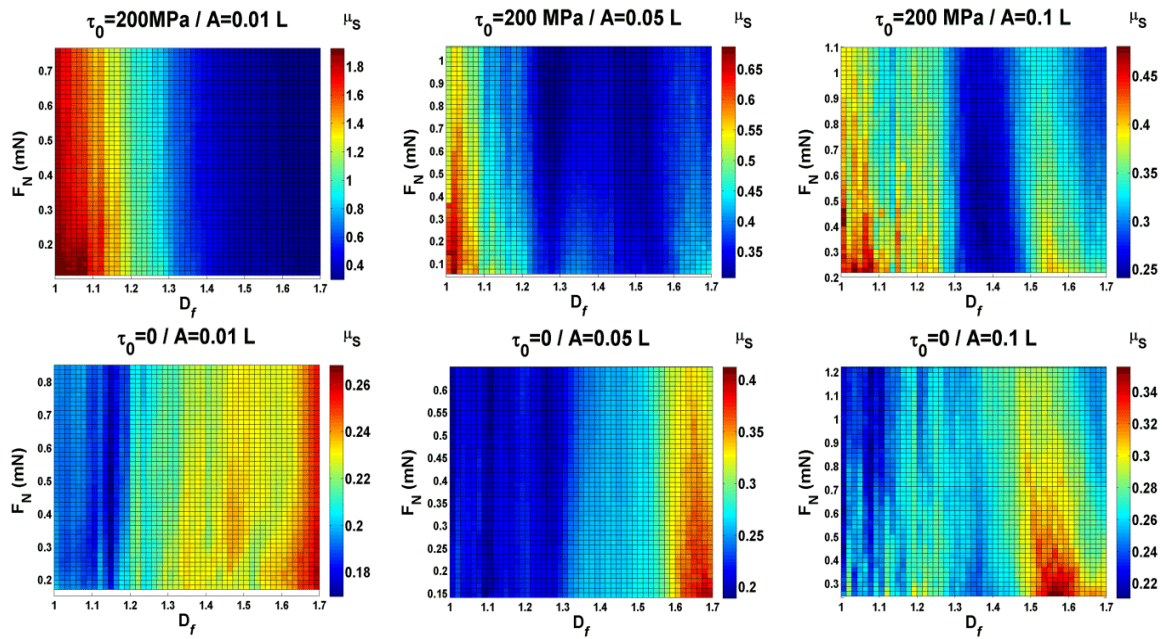

**Figure 7. Variation of macroscopic friction coefficient ($\mu_s$, colour bars) with fractal dimension and applied normal load for different conditions of asperity aspect ratio (A) and adhesive strength ($\tau_0$).**

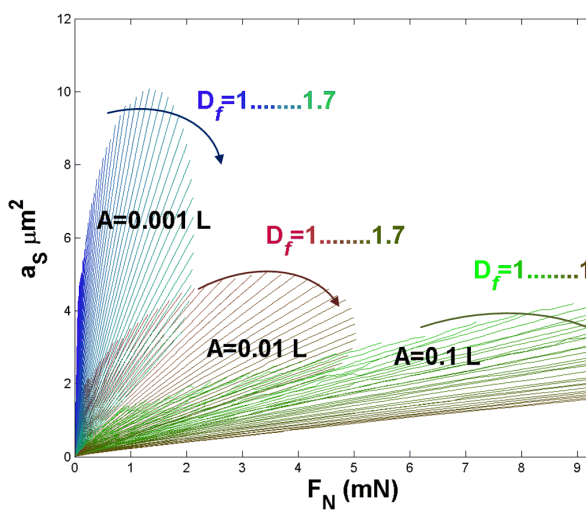

**Figure 8. Variation of true contact area ($a_s$) with applied normal load ($F_N$) for surfaces of different fractality across three data sets differed by their relative amplitudes (A).**

The linearity between true contact area and applied load for surfaces of varied fractality is shown in Figures 5 and 8 and, as a commonly utilised benchmark, serves to validate the methods applied here. Although, it is likely that frameworks incorporating elasto-plastic asperity deformation would yield a more meaningful interpretation with respect to engineering materials [65]. Importantly, the SAAD method facilitates the study of contact mechanics and friction interactions in multiscale surfaces that would otherwise necessitate computationally intensive analysis. For given values of $\tau_0$ and $\mu_0$ and $D_f$ the repeated analysis of contact events between two surface profiles consisting each of $1.5 \times 10^5$ points, over 100 reiterations was achieved in the order of one minute. This computationally efficient approach thus facilitates parametric studies including of a broad range of variables here chosen as fractality, adhesive strength, molecular friction and surface amplitude. The execution of such a methodology in a finite element framework would be highly problematic owing to the large number of elements that would be necessary in order to mesh the structures adequately with statistically significant surface generation. Alternative approaches to the computational description of interfaces have utilised principal component analysis, Fourier transforms or other mathematical formulations to represent periodic multiscale surface profiles [98]. However, this approach, while enabling computationally efficient parametric studies, may not meaningfully capture the nature of real surfaces in terms of random fractal asperity structures.

Results demonstrate the interplay between parameters of applied load, surface fractality and molecular mechanisms of friction. These relationships are evident of the necessity to consider cross scale surface structures and surface chemistry in tandem with loading conditions to evaluate the development of frictional interactions, which for certain systems may indeed exhibit a non-monotonous variation with surface fractality. The results of the presently applied methods have significance towards understanding the evolution of force networks in multi body systems and under conditions of low normal load, where non-linearity between normal load and frictional forces may be observed, and evolving surface structure and interface chemistry can fundamentally alter bulk system behaviour. Numerous simplifications are involved in the present work, such as an atomic/molecular friction coefficient that is independent of load and material orientation, where in fact friction at atomistically flat interfaces is generally reported to vary with load and crystallographic orientation of the surfaces in contact, with negative friction occurring under certain conditions [99].

On the basis of the present method, for surfaces of greater fractality, we predict a higher macroscopically observed friction coefficient, with this trend diminishing for surface conditions that are conducive to bonding and the formation of adhered junctions. This trends are consistent with behaviour experimentally observed in certain systems [100]. Additionally, an increase in lowest scale friction occurring through molecular interactions, as represented by $\mu_0$, produces an increase in the magnitude of macroscopic friction for all surface structures studied, with this trend expected to be somewhat greater for surfaces of greater fractality. Lower asperity aspect ratios, which generally are found following polishing treatments, naturally give rise to a higher true contact area under a given applied load, increasing frictional forces for surface of lower fractality and leading to greater significance of potential adhesive interactions.

It is worth noting that at rough interfaces the magnitude of load-independent bonding strength may paradoxically depend on the nature of loading history. This is often the case for rough metallic interfaces where a high asperity-localised load may facilitate metal-metal bonding while oxide or hydration-passivated layers exhibit weak van der Waals mediated interactions. For this reason the assumption that the value of $\tau_0$ is constant at all contact regions merits revision in future work.

## 5. Conclusions

We have developed and demonstrated the application of a computationally efficient method for the study of rough to rough surface interactions towards the prediction of static frictional forces on the basis of discretisation of a hierarchical surface structure. The present method involves simplifications in terms of surface mechanics and phyisco-chemical surface interactions but nonetheless facilitates the interpretation of the dependence of frictional forces on parameters of surface fractality, molecular scale friction and adhesive type interactions. Results show that under certain conditions friction may be minimised at an intermediate regime of surface fractality and further confirm that the macroscopically observed friction coefficient for a rough surface in a particular system can be lower than the friction coefficient of an atomically smooth surface of the same material.

In conjunction with the characterisation of specific material surface interactions and the evaluation of multi-scale surface structures, the methods developed have the potential to inform the modification of surface structures towards optimisation of frictional interactions and interpretation of mechanical phenomena in multi-body and micro scale systems.

## Acknowledgements:

Financial support for this research from the Australian Research Council through grant no. DP120104926 is gratefully appreciated.